\documentclass[prl,twocolumn]{revtex4}

\usepackage{amsmath}
\usepackage[english]{babel}

\begin{document}

\title{Holonomic quantum computation in subsystems}
\author{Ognyan Oreshkov} \affiliation{Grup de F\'{i}sica Te\`{o}rica, Universitat Aut\`{o}noma de
Barcelona, 08193 Bellaterra (Barcelona), Spain}

\begin{abstract}
We introduce a generalized method of holonomic quantum computation
(HQC) based on encoding in subsystems. As an application, we propose
a scheme for applying holonomic gates to unencoded qubits by the use
of a noisy ancillary qubit. This scheme does not require
initialization in a subspace since all dynamical effects factor out
as a transformation on the ancilla. We use this approach to show how
fault-tolerant HQC can be realized via 2-local Hamiltonians with
perturbative gadgets.
\end{abstract}

\pacs{03.67.Pp, 03.65.Vf}
\maketitle


\textit{Introduction}.---A basic requirement for the construction of
a reliable quantum computer is the ability for high-fidelity storage
and manipulation of quantum information. Storage is most generally
achieved through encoding in suitable physical degrees of freedom
that may be protected from unwanted interactions through symmetries
or by active operations. At the same time, manipulating quantum
information requires addressing the relevant degrees of freedom in a
precise and robust way.

Holonomic quantum computation (HQC) is one method that promises a
resilient way of information processing through all-geometric,
adiabatic control \cite{HQC}. In this approach, logical states are
encoded in a degenerate eigenspace of the Hamiltonian, and gates are
realized by adiabatically varying the Hamiltonian along suitable
paths in the space of control parameters. This gives rise to
geometric transformations inside the eigenspace \cite{WZ84}, which
depend on certain global properties of the path and are thus robust
against local fluctuations that preserve those properties
\cite{HQCrob}. This method employs encoding in a subspace, which is
only a special case of the most general form of encoding possible
\cite{Knill06}---encoding in \textit{virtual} subsystems \cite{vs}.
The latter type of encoding has numerous applications in the area of
decoherence control, ranging from noiseless subsystems that offer
protection through more efficient encoding or in cases where no
subspace protection exists \cite{DFs}, to operator error-correcting
codes \cite{OQEC} that allow for simplified recovery methods leading
to improved fault-tolerance thresholds \cite{AC07}. Given the
operational robustness of the holonomic approach, a natural question
is whether a generalized method for HQC compatible with encoding in
subsystems is possible.

In this paper, we answer the above question affirmatively. The paper
reports three main results. The first one is a general framework for
HQC in subsystems. A distinctive feature of this framework is that
it involves performing simultaneous computations in the different
eigenspaces of the Hamiltonian. We show that given sufficient
control over the parameters of a Hamiltonian, it is possible to
generate any combination of geometric transformations in its
eigenspaces. This possibility has been suggested in Refs.~\cite{vs,
ZanardiLloyd}, but a proof has been lacking. A remarkable
consequence of this result is the possibility to apply purely
geometric operations on a given system by the use of a noisy
ancilla. This approach does not require preparation of the system in
any subspace since all dynamical effects are absorbed by the
ancilla. The latter is particularly appealing as it avoids the
problem of imperfect initialization \cite{ini}. The second main
result is a scheme for universal quantum computation on qubits based
on this principle. The scheme offers a robust way of gate
implementation which in comparison to standard HQC is less demanding
in the preparation stage and insensitive to certain transitions
between energy levels.

The third main result is a scheme for fault-tolerant (FT) HQC based
on 2-local interactions. The ultimate scalability of any method of
computations requires the ability for FT implementation that
guarantees the existence of an accuracy threshold \cite{FT}. The
first such HQC scheme \cite{OBL08} requires 3-local interactions
that cannot be reduced to 2-local using standard techniques without
losing fault tolerance. As 3-local interactions can be very
difficult to engineer, the question of whether FTHQC can be
implemented using 2-local Hamiltonians is especially important in
view of possible experimental implementations. We show that our
scheme for HQC with noisy ancillas is readily compatible with the FT
techniques on stabilizer codes \cite{FT} and can be reduced to
2-local by perturbative gadgets \cite{gadgets} in a FT way.

\textit{Quantum holonomies}.---Consider an isodegenerate family of
Hamiltonians  $\{H_{\lambda}\}$ on an $N$-dimensional Hilbert space,
continuously parameterized by a parameter $\lambda$ in a
control-parameter manifold $\mathcal{M}$: $H_{\lambda}=\sum_{n=1}^{R}%
\varepsilon_n(\lambda)\Pi_n(\lambda)$, where $\{\varepsilon_n(\lambda)%
\}_{n=1}^{R}$ are the $R$ different $d_n$-fold degenerate eigenvalues of $%
H_\lambda$, ($\sum_{n=1}^{R} d_n=N$), and $\Pi_n(\lambda)$ are the
projectors on the corresponding eigenspaces. Let $\lambda^\mu$
be local coordinates on $\mathcal{M}$ ($1\leq\mu\leq \text{dim}\mathcal{M}$%
) and $\{|n\alpha; \lambda\rangle\}_{\alpha=1}^{d_n}$ be an
orthonormal basis of the $n^{\text{th}}$ eigenspace of the Hamiltonian at the point $%
\lambda$. Consider a time-dependent Hamiltonian
$H(t):=H_{\lambda(t)}$ obtained by varying the control parameters
along a curve $\lambda(t)$ in $\mathcal{M}$. If the change of
$\lambda$ is adiabatic, the transformation generated by $H(t)$ is
$U(t)=\mathcal{T}\text{exp}(-i\int_0^t d\tau H(\tau)) =
\oplus_{n=1}^R e^{i\omega_n(t)}U^{\lambda}_{A_n}(t)$, where
$\omega_n(t)=-\int_0^td\tau\varepsilon_n(\lambda(\tau))$ are
dynamical phases, and $U^{\lambda}_{A_n}(t)=\mathcal{P}\text{exp}(\int_{\lambda(0)}^{%
\lambda(t)}A_n)$, where $\mathcal{P}$ denotes path-ordering. The adiabatic connections are $%
A_n=\sum_\mu A_{n,\mu} d\lambda^\mu$, where $A_{n,\mu}$ has matrix
elements
$(A_{n,\mu})_{\alpha\beta}=\langle n\alpha; \lambda|\frac{%
\partial}{\partial \lambda^\mu}|n\beta;\lambda\rangle$ \cite{WZ84}.
When the path $\lambda (t)$ forms a loop $\gamma (t)$, $\gamma
(0)=\gamma (T)=\lambda _{0}$, the unitary matrix $U_{n}^{\gamma
}\equiv U_{A_{n}}^{\lambda }(T)=\mathcal{P}\text{exp}(\oint_{\gamma
}A_{n})$ is
called the holonomy associated with the loop. The set $\text{Hol}%
(A_{n})=\{U_n^{\gamma }|\gamma \in L_{\lambda _{0}}(\mathcal{M})\}$,
where $L_{\lambda _{0}}(\mathcal{M})=\{\gamma :[0,T]\rightarrow \mathcal{M}%
|\gamma (0)=\gamma (T)=\lambda _{0}\}$ is the space of all loops
based on $\lambda _{0}$, is a subgroup of $U(d_{n})$ called the
holonomy group. In Ref.~\cite{HQC} it was argued that in the generic
case, the adiabatic connection corresponding to the
$n^{\textrm{th}}$ energy level of the Hamiltonian family is
irreducible, i.e., the holonomy group $\text{Hol}%
(A_{n})$ is equal to $U(d_{n})$, and therefore adiabatic holonomies
can be used for universal quantum computation in the
$n^{\textrm{th}}$ eigenspace of the Hamiltonian.

\textit{HQC in subsystems}.---A (virtual) subsystem \cite{vs} is any
tensor factor of a subspace of a system's Hilbert space
$\mathcal{H}^S$. We will consider decompositions of the form
\begin{equation}
\mathcal{H^S}=\bigoplus_{i=1}^m\mathcal{H}^A_i\otimes\mathcal{H}^B_i,\label{decomposition}
\end{equation}
where logical states are encoded in the subsystems
$\mathcal{H}^A_i$, and will be interested in universal HQC in
$\{\mathcal{H}^A_i\}$. Eq.~\eqref{decomposition} describes the most
general form of encoding of information and plays a fundamental role
in the theory of quantum error correction \cite{OQEC} where it
provides the structure of preserved information under open-system
dynamics \cite{KNPV07}. (Note that most generally all subsystems
$\mathcal{H}^A_i$ can be used for encoding and computation
simultaneously \cite{BKK07}).

\textit{Lemma}. Let $H(0)$ be a Hamiltonian with at least two
different eigenvalues. It is possible to implement any combination
of holonomies in the eigenspaces of $H(0)$ through a suitable
adiabatic cyclic change of $H(0)$.

\textit{Comment}. It is known that given sufficient control over the
parameters of a Hamiltonian we can generate holonomically any
unitary operation in a given eigenspace \cite{HQC}. The question we
address here is whether it is possible to generate an arbitrary
\textit{combination} of holonomies in the different eigenspaces.
This property may not be obvious. For example, in the case of a
two-level Hamiltonian, the evolution of one eigenspace completely
determines the evolution of the other one. Since the holonomy in a
given eigenspace depends entirely on the evolution of that
eigenspace, this raises the question if the two eigenspaces can
undergo arbitrary independent holonomies. We now show, by
construction, that this is possible.

\textit{Proof}. It is sufficient to show that it is possible to
generate a universal set of gates in any given eigenspace while at
the same time generating the identity operation in the rest of the
eigenspaces. Without loss of generality, we will assume that there
are only two eigenspaces; if there are more, we can always operate
within the subspace spanned by two of them by varying only the
restriction of the Hamiltonian on that subspace. Then the initial
Hamiltonian can be written
$H(0)=\varepsilon_1\Pi_1+\varepsilon_2\Pi_2$, where $\Pi_{1,2}$ are
the projectors on the ground and excited eigenspaces, and
$\varepsilon_1<\varepsilon_2$ are their corresponding eigenvalues.
Observe that $H(0)$ is invariant under unitaries of the form
$V=V_1\oplus V_2,\label{V}$ where $V_{1,2}$ are unitaries on the
subspaces with projectors $\Pi_{1,2}$, respectively. Let the
Hamiltonian vary along a loop $H(t)$, $H(0)=H(T)$, which satisfies
the adiabatic requirement to some satisfactory precision. To this
precision, the resulting unitary transformation can be written
$U(T)=\mathcal{T}\textrm{exp}(-i\int_0^TdtH(t))=e^{-i\omega_1}U_1
\oplus e^{-i\omega_2}U_2$, where $U_{1}$ and $U_2$ are the
holonomies resulting in the two eigenspaces, and
$\omega_{1,2}=\int_0^T dt\varepsilon_{1,2}(t)$ are dynamical phases.
Observe that the Hamiltonian $VH(t)V^{\dagger}$, where $V=V_1\oplus
V_2,\label{V}$ gives rise to the holonomies $V_1U_1V_1^{\dagger}$
and $V_2U_2V_2^{\dagger}$, respectively. This follows from the fact
that the overall unitary transformation generated by
$VH(t)V^{\dagger}$ is equal to $VU(t)V^{\dagger}$ where $U(t)$ is
the unitary generated by $H(t)$. Note that $VH(t)V^{\dagger}$ is a
valid loop based on $H(0)$ with the same spectrum as that of $H(t)$.

Imagine that we want to generate holonomically a gate $W_1$ in the
ground space of the Hamiltonian while at the same time realizing the
identity holonomy $I_2$ in the excited space. Choose any loop $H(t)$
which gives rise to the holonomy $W_1^{\frac{1}{d_2}}$ in the ground
space, where $d_2$ is the dimension of the excited space (we know
that such a loop can be found). Let this loop result in the holonomy
$W_2$ in the excited space. The latter can be written
$W_2=\overset{d_2}{\underset{j=1}{\sum}}
e^{i\alpha_j}|j\rangle\langle j|$, where $\{|j\rangle\}$ is an
eigenbasis of $W_2$ and $e^{i\alpha_j}, \alpha_j\in R$, are the
corresponding eigenvalues. Consider the unitary $C_2$ which cyclicly
permutes the eigenvectors $\{|j\rangle\}$:
$C_2|j\rangle=|j+1\rangle$, where we define $|d_2+1\rangle\equiv
|1\rangle$. We can implement the desired combination of holonomies
in the two eigenspaces as follows. First apply $H(t)$. This results
in the holonomies $W_1^{\frac{1}{d_2}}$ and $W_2$ in the ground and
excited spaces, respectively. Next, apply $(I_1\oplus
C_2)H(t)(I_1\oplus C_2)^{\dagger}$. This generates the holonomies
$W_1^{\frac{1}{d_2}}$ and
$C_2W_2C_2^{\dagger}=\overset{d_2}{\underset{j=1}{\sum}}
e^{i\alpha_{j-1}}|j\rangle\langle j|$ (we have defined
$\alpha_{1-1}\equiv \alpha_{d_2}$). The combined effect of these two
operations is $W_1^{\frac{2}{d_2}}$ and
$\overset{d_2}{\underset{j=1}{\sum}}
e^{i(\alpha_j+\alpha_{j-1})}|j\rangle\langle j| $. We next apply
$(I_1\oplus C_2^2)H(t)(I_1\oplus C_2^2)^{\dagger}$, which generates
the holonomies $W_1^{\frac{1}{d_2}}$ and
$C_2^2W_2C_2^{2\dagger}=\overset{d_2}{\underset{j=1}{\sum}}
e^{i\alpha_{j-2}}|j\rangle\langle j|$. The net result becomes
$W_1^{\frac{3}{d_2}}$ and $\overset{d_2}{\underset{j=1}{\sum}}
e^{i(\alpha_j+\alpha_{j-1}+\alpha_{j-2})}|j\rangle\langle j| $. We
continue this for a total of $d_2$ rounds, which results in the net
holonomic transformations $W_1^{\frac{d_2}{d_2}}=W_1$ and
$e^{i(\alpha_1+\alpha_2+...+\alpha_{d_2})}
\overset{d_2}{\underset{j=1}{\sum}} |j\rangle\langle j|\propto I_2$.
This completes the proof.

The proof uses sequences of loops. In the next section, we will see
that depending on the task it may be possible to find constructions
based on a single loop.

\textit{Theorem 1}. Consider a nontrivial subsystem decomposition of
the form \eqref{decomposition}. Choose an initial Hamiltonian in the
form $ H(0)=\bigoplus_{i=1}^m I^A_i\otimes H^B_i$, where $H^B_i$ are
operators on $\mathcal{H}^B_i$ such that all eigenvalues of $H^B_i$
are different from the eigenvalues of $H^B_j$ for $i\neq j$. In the
case when $m=1$, we impose the additional requirement that $H^B_1$
has at least two different eigenvalues. By varying adiabatically
this Hamiltonian along suitable loops in a sufficiently large
control manifold, it is possible to generate a unitary of the form
$U=\bigoplus_iW_i^A\otimes V^B_i$, where $\{W_i^A\}$ is an arbitrary
set of geometric transformations on $\{\mathcal{H}^A_i\}$.

\textit{Proof}. Denote the eigenvalues of $H^B_i$ by
$\omega_{\alpha_i}$, $\alpha_i=1,...,d_i$, and the projectors on
their corresponding eigenspaces $\mathcal{H}^B_{\alpha_i}$ by
$\Pi^B_{\alpha_i}$. Then $H(0)$ has the spectral decomposition
$H(0)=\sum_{i=1}^m\sum_{\alpha_i=1}^{d_i}\omega_{\alpha_i}
\Pi^A_{i}\otimes \Pi^B_{\alpha_i}$, where $\Pi^A_{i}$ is the
projector on $\mathcal{H}^A_i$. According to Lemma 1, we can
implement holonomically any combination of unitary transformations
in the different eigenspaces of $H(0)$ up to an overall phase. Thus
by applying the holonomy $W^A_i\otimes W^B_{\alpha_i}$ in each of
the eigenspaces $\mathcal{H}^A_i\otimes\mathcal{H}^B_{\alpha_i}$ for
$\alpha_i=1,...,d_i$, where $W^B_{\alpha_i}$ are arbitrary unitaries
on $\mathcal{H}^B_{\alpha_i}$, we obtain the net unitary
$U=\bigoplus_iW_i^A\otimes
(\bigoplus_{\alpha_i}e^{i\phi_{\alpha_i}}W^B_{\alpha_i})\equiv
\bigoplus_iW_i^A\otimes V^B_i$, where $e^{i\phi_{\alpha_i}}$ are
dynamical phases.

\textit{HQC without initialization}.---From Theorem 1 one can see
that in the case when the Hilbert space factors as
$\mathcal{H}^S=\mathcal{H}^A\otimes\mathcal{H}^B$, it is possible to
apply holonomic computation in subsystem $\mathcal{H}^A$ without
initializing the state of the system in any subspace. In particular,
if we are given a system $\mathcal{H}^A$ in an unknown state, we can
append to it another ancillary system $\mathcal{H}^B$, also in an
unknown state, and apply any desired transformation holonomically on
the first system. Since this approach does not require the
preparation of pure ancillary states, it can be advantageous in
implementations where the latter is difficult, such as nuclear
magnetic resonance (NMR) \cite{NMR}.

We now present an explicit scheme for universal computation on
qubits based on this principle. The scheme uses a single ancillary
gauge qubit. We will show how to implement a universal set of one-
and two-qubit gates. We will build the necessary loops by
interpolations between points in the space of Hamiltonians of the
form $H(t)=f(t)H(0)+g(t)H(T)$, where $f(0)=g(T)=1$, $f(T)=g(0)=0$.
This Hamiltonian interpolates between $H(0)$ and $H(T)$ during a
time interval $T$. The interpolating Hamiltonians that we will be
using (to be described below) have two energy levels of equal
degeneracy, and their energy gaps are non-zero unless the entire
Hamiltonian vanishes, i.e., the interpolations can be realized
adiabatically for sufficiently long time $T$ and smooth choice of
$f(t)$ and $g(t)$. For an adiabatic interpolation of the above type
we will use the short notation $H(0)\rightarrow H(T)$.

Let us label the two qubits on which we will be applying the gates
by 1 and 2, and the gauge qubit by 3. In order to apply a
single-qubit gate, say, on qubit 2, we will use the starting
Hamiltonian $H(0)=I_2\otimes X_3$. (Here by $X_i$, $Y_i$, $Z_i$ and
$I_i$ we denote the Pauli matrices and the Identity acting on the
$i^{\textrm{th}}$ qubit). We first apply
\begin{equation}
I_2\otimes X_3 \rightarrow Z_2\otimes Z_3.\label{IXZZ}
\end{equation}
This results in a geometric transformation and a dynamical phase in
each eigenspace. Let us denote the purely geometric part of the
resulting unitary by $U_{2,3}$ (the exact form of $U_{2,3}$ is not
important since we will undo it later). At this point we can apply a
unitary whose geometric part is equal to an arbitrary gate on qubit
2 according to a method described in Ref.~\cite{OBLprep}. For
example, the interpolation $Z_2\otimes Z_3\rightarrow X_2\otimes
Z_3$ gives rise to the geometric operation $R_2Z_2$, where $R_2$ is
the Hadamard gate on qubit 2. The interpolation $Z_2\otimes Z_3
\rightarrow -(\cos{\frac{\pi}{8}}X_2+\sin{\frac{\pi}{8}}Y_2)\otimes
Z_3\rightarrow -Z_2\otimes Z_3$ results in the geometric operation
$T_2Z_2X_2$ , where $T_2$ denotes the $\pi/8$ gate on qubit 2. These
two gates are a universal set of single-qubit gates. Let $G_2$ be
the gate from the above set which we want to implement. After the
corresponding interpolation, the net geometric transformation
becomes $G_2U_{2,3}$ and the Hamiltonian is transformed to
$G_2Z_2G_2^{\dagger}\otimes Z_3$. We can now ``undo" the unitary
$U_{2,3}$ by applying the interpolation
$G_2Z_{2}G_{2}^{\dagger}\otimes Z_3\rightarrow I_{2}\otimes X_3$.
The latter is the inverse of Eq.~\eqref{IXZZ} up to the single-qubit
unitary transformation $G_2$, i.e., it results in the transformation
$G_2U_{2,3}^{\dagger}G_{2}^{\dagger}$. Thus the net result is $
G_2U_{2,3}^{\dagger}G_2^{\dagger}G_2U_{2,3}=G_2$, which is the
desired unitary on qubit $2$. Note that the relative dynamic phase
between the ground and excited spaces, which accumulates during the
procedure, at the end is equivalent to a transformation on qubit 3.

For universal computation, we also need a nontrivial two-qubit gate.
We can start again by the interpolation \eqref{IXZZ} which results
in the geometric transformation $U_{2,3}$. At this point we can
apply, for example, the interpolation $I_1\otimes Z_2\otimes
Z_3\rightarrow I_1\otimes Y_2\otimes Z_3 \rightarrow Z_1\otimes
Z_2\otimes Z_3$, which results in the gate $S_1^{\dagger}N_{1,2}$,
where $S_1=T_1^2$ and $N_{1,2}$ is the ``controlled not'' gate with
qubit 1 the control, and qubit 2 the target \cite{OBLprep}. To
``undo" the operation $U_{2,3}$, we apply the transformation $
Z_1\otimes Z_2\otimes Z_3 \rightarrow I_1\otimes I_2\otimes X_3$,
which is the inverse of \eqref{IXZZ} up to the transformation
$S_1^{\dagger}N_{1,2}$. The net result is
$S_1^{\dagger}N_{1,2}U^{\dagger}_{2,3}N_{1,2}S_1S_1^{\dagger}N_{1,2}U_{2,3}=S_1^{\dagger}N_{1,2}$.

We note that unlike the standard holonomic approach, here each
eigenspace of the Hamiltonian undergoes the same geometric
operation, which supplies the scheme with additional robustness. The
scheme is insensitive to those transitions between the two energy
levels that are equivalent to local operations on the transformed
gauge qubit. The scheme uses 2- and 3-local Hamiltonians.

\textit{FTHQC with 2-local Hamiltonians}.---The theory of fault
tolerance \cite{FT} guarantees that, if errors during the
implementation of a given gate are sufficiently uncorrelated and
improbable, an arbitrarily long computation can be implemented
reliably with a modest resource overhead. The first proposal for
FTHQC \cite{OBL08} uses the encoding present in a stabilizer code
and Hamiltonians that are elements of the instantaneous stabilizer
or gauge group of the code. These Hamiltonians couple qubits in the
same block, but errors do not propagate as each eigenspace is
subject to the same transversal operation. That scheme requires
3-local Hamiltonians. Even though every Hamiltonian can be simulated
by a 2-local one via the so called \textit{perturbative gadgets}
\cite{gadgets}, the locality of that scheme cannot be reduced by a
direct application of these techniques. Since the simulated
Hamiltonian couples qubits in the same code block, an error on one
of the gadget ancillas \cite{gadgets} can spread to multiple qubits
within a block.

The scheme from the previous section suggests an alternative
approach to FTHQC. Any FT protocol on qubit stabilizer codes can be
decomposed into transversal one- and two-qubit gates (these are
gates that couple only the corresponding qubits from different
blocks). In addition, one requires the preparation of a special
ancillary state such as $(|00...0\rangle+|11...1\rangle)/\sqrt{2}$,
which is done non-transversally. Transversality guarantees that a
single error during an encoded operation results in at most one
error per block of the code. Our scheme for holonomic one- and
two-qubit gates is readily compatible with this approach: we can
apply the same operations as in a standard FT protocol \cite{FT} by
coupling every qubit or pair of qubits in the code to an ``external"
gauge qubit as described in the previous section. Obviously, a
single error during the implementation of a transversal operation
cannot propagate to multiple qubits in a block because the latter do
not interact. In contrast to the previous approach which can be
understood as performing HQC inside the subsystem containing the
protected information, this scheme performs HQC in the entire system
and does not require Hamiltonians that depend on the code.
Furthermore, here each 3-local Hamiltonian can be reduced to 2-local
via perturbative gadgets as the gadget ancillas would couple to at
most one qubit inside a block. A complete fault-tolerance analysis
is beyond the scope of this paper, but we note that the use of extra
qubits increases the chance for an error during a single gate. In
addition, the 3-qubit gadget decreases the minimum gap of the
original Hamiltonian by a factor $\sim\varepsilon^{-3}$ where
$\varepsilon$ is the perturbation parameter in which the
approximation to order $O(\varepsilon^4)$ is carried out
\cite{gadgets}. Thus for maintaining a given precision, the time for
implementing a two-qubit gate would have to increase by a similar
factor which decreases the allowed rate for environment noise. A way
around this could be to look for non-perturbative implementations.
Since the no-initialization property is not crucial for fault
tolerance \textit{per se}, implementations with an ancilla in a
known state can also be considered.

\textit{Conclusion}.---In summary, we have introduced a general
framework for HQC in subsystems, showing that it is possible to
realize simultaneously independant HQC in the subsystems
$\mathcal{H}^A_i$ in any nontrivial decomposition of the form
\eqref{decomposition}. As an application, we proposed a robust
scheme for applying purely geometric gates to unencoded qubits by
the use of a noisy ancillary qubit. We used this approach to show
that 2-qubit Hamiltonians are universal for FTHQC. We hope that our
results will open new avenues for quantum information processing
implementations that combine the robustness of the holonomic control
with the most general form of encoding. An interesting future
direction would be to extend the present results to the theory of
geometric phases based on dynamical invariants \cite{Duzz08}, which
encompasses non-adiabatic, mixed-state and open-path holonomies.
Since geometric phases have wide applications, associating
holonomies with subsystems could find use beyond the field of
quantum computing as well.

\textit{Acknowledgements}.---OO was supported by Spanish MICINN
(Consolider-Ingenio QOIT). The author thanks D. Poulin for
suggesting the gadget approach, and P. Zanardi, D. A. Lidar, T. A.
Brun, and J. Calsamiglia for helpful discussions.

\textit{Note}.---Recently, an independent scheme for (open-path)
FTHQC with 2-local Hamiltonians via perturbative gadgets was
proposed by D. Bacon and S. T. Flammia in Ref.~\cite{BF09}.

\end{document}